\documentclass[11pt,a4paper]{article}
\pdfoutput=1
\usepackage{jheppub}

\newcommand{\as}{\alpha_s}
\newcommand{\wh}{\widehat}
\newcommand{\nn}{\nonumber}
\newcommand{\ol}{\overline}

\newcommand{\po}{\phantom{00}}
\newcommand{\eqn}[1]{(\ref{#1})}

\newcommand{\tvs}{\vbox{\vskip 6mm}}
\newcommand{\mvs}{\vbox{\vskip 8mm}}

\newcommand{\sfrac}[2]{\mbox{$\frac{#1}{#2}$}}

\title{Anomalous dimensions of four-quark operators and renormalon structure
       of mesonic two-point correlators}

\author[a,b]{Diogo Boito,}
\author[c]{Dirk Hornung}
\author[d]{and Matthias Jamin}

\affiliation[a]{Instituto de F\'isica de S\~ao Carlos, Universidade de
                S\~ao Paulo,\\ CP 369, 13560-970, S\~ao Carlos, SP, Brazil}
\affiliation[b]{Instituto de F\'\i sica, Universidade de S\~ao Paulo,
                Rua do Mat\~ao Travessa R, 187,\\ 05508-090, S\~ao Paulo,
                SP, Brazil}
\affiliation[c]{Institut de F\'\i sica d’Altes Energies (IFAE), The Barcelona 
                Institute of Science and Technology,\\ Campus UAB,
                08193 Bellaterra (Barcelona) Spain}
\affiliation[d]{Instituci\'o Catalana de Recerca i Estudis Avan\c{c}ats
                (ICREA),\\
                Institut de F\'\i sica d’Altes Energies (IFAE), The Barcelona
                Institute of Science and Technology,\\ Campus UAB,
                08193 Bellaterra (Barcelona) Spain}

\emailAdd{boito@ifsc.usp.br}
\emailAdd{dirk.hornung@uab.cat}
\emailAdd{jamin@ifae.es}

\abstract{In this work, we calculate leading-order anomalous dimension matrices
for dimension-6 four-quark operators  which appear in the operator product
expansion of flavour non-diagonal and diagonal vector and axial-vector two-point
correlation functions. The infrared renormalon structure corresponding to
four-quark operators is reviewed and it is investigated how the eigenvalues of
the anomalous dimension matrices influence the singular behaviour of the $u=3$
infrared renormalon pole. It is found that compared to the large-$\beta_0$
approximation where at most quadratic poles are present, in full QCD at $N_f=3$
the most singular pole is more than cubic with an exponent $\kappa\approx 3.2$.
}%


\begin{document}
\maketitle

\section{Introduction}\label{sect1}

The perturbative expansion in QCD is known to lead to a divergent series
which is at best asymptotic. The asymptotic behaviour of the perturbative
series manifests itself in the appearance of singularities for its Borel
transform which lie on the negative or positive real axis in the Borel
variable. Those singularities, connected with renormalisation of the theory,
are termed {\em renormalons} \cite{tHo79,ben98}. More specifically,
singularities related to the short-distance behaviour of the theory lie on the
negative real axis and are called ultraviolet (UV) renormalons. Those related
to long-distance physics appear on the positive real Borel axis and are termed
infrared (IR) renormalons.

The presence of IR renormalon poles leads to ambiguities in the definition of
the full function which is related to the perturbative series, because the
Borel resummation (inverse Borel transform) entails to perform an integral
over the positive real Borel axis which naively is not well defined. The
ambiguities in the definition of the Borel integral are exponentially small
terms in the QCD coupling, $\alpha_s$. Associated with them is the appearance
of higher-dimensional operator corrections, the so-called QCD {\em condensates},
such that the full function is unambiguous. The operators that display
renormalon ambiguities are a subset of those that arise in the framework of
the operator product expansion (OPE). 

The renormalisation group (RG) allows one to relate the exponent of a given
renormalon pole to the leading-order anomalous dimension of the associated
operator. Therefore, knowledge about the anomalous dimensions of operators
entering the OPE expansion of a correlator can be translated into knowledge of
the Borel transform of its purely perturbative contribution. The position and
strength of poles can, in principle, be predicted in this way while the residua,
which are non-perturbative, cannot. This partial knowledge of the Borel
transformed perturbative series can be exploited as a way to gain understanding
about higher orders, yet unknown from loop computations. In particular, in the
case of the QCD description of hadronic $\tau$ decays~\cite{bnp92}, models of
the Borel transform have been used in order to assess -- among other things --
the virtues of different RG improvement prescriptions~\cite{bj08,bbj13}. It
is with this type of application in mind that we revisit the computation of
anomalous dimensions of four-quark operators.

Limiting ourselves to correlation functions of vector or axial-vector currents
with respect to the QCD vacuum, the IR renormalon pole on the positive real
axis closest to the origin of the Borel plane is associated to the vacuum
matrix element of one dimension-four operator, the {\em gluon condensate}. The
next-to-closest singularity then is found to correspond to the dimension-6
{\em  triple gluon condensate} and a set of dimension-6 {\em four-quark
condensates}. It is these latter four-quark condensates that we investigate in
more detail in the present work.

The central aim of the present study is to provide the leading-order anomalous
dimension matrices corresponding to the dimension-6 four-quark operators that
emerge in the OPE of two-point correlation functions of flavour non-diagonal,
as well as flavour diagonal, mesonic vector and axial-vector currents. Those
anomalous dimensions contain information about the structure of the related
IR renormalon poles. The computation of leading-order anomalous dimensions of
four-fermion operators is fairly standard \cite{mm83}, and in ref.~\cite{jk86},
results were presented for a complete set of spin-0 four-quark operators
without derivatives in the case of three active light quark flavours, $N_f=3$.

As a matter of principle, all findings presented in this work are derivable
from the results of ref.~\cite{jk86} through operator relations, valid in
four space-time dimensions. However, the results in \cite{jk86} (and also
\cite{mm83}), were only given for a number of quark colours $N_c=3$, and here
we intend to provide results at arbitrary $N_c$. Furthermore, in order to be
able to connect to the large-$N_f$, or the related large-$\beta_0$ limit, of
QCD \cite{ben98}, in contrast to ref.~\cite{jk86}, explicit $N_f$ dependencies
will be kept. To this end, we therefore followed two alternative routes: first,
to recalculate ref.~\cite{jk86} at arbitrary $N_c$, which is discussed in
appendix~\ref{appA}, and then derive the results below from the mentioned
operator relations. And second, to compute the anomalous dimensions directly
for the operators appearing in the correlation functions. Both approaches lead
to agreeing results and also at $N_c=N_f=3$, the anomalous dimension matrices
of appendix~\ref{appA} coincide with \cite{jk86}. Furthermore, in some test
cases that we checked, also full agreement with the results of ref.~\cite{mm83}
was found.

In summary, the material presented in this work is organised as follows: first,
in section~\ref{sect2}, we review the next-to-leading order results on the
dimension-6 four-quark operator contributions to the flavour non-diagonal
vector and axial-vector correlation functions available in the literature
\cite{ac94,lsc86}. For the set of appearing operators, we calculate the
leading-order anomalous dimension matrices. In order to obtain a set which
closes under renormalisation, three dimension-6 four-quark operators of
penguin type have to be included. Analogous results are provided in
section~\ref{sect3} for the flavour diagonal vector and axial-vector
correlators.

Next, we construct the operator combinations for which the leading-order 
anomalous dimension matrices are diagonal and compute the corresponding
eigenvalues. These eigenvalues are ingredients for the renormalon structure
of the perturbative series which is related to the four-quark operators and
discussed in section~\ref{sect4}. Along these lines, we determine the singular
behaviour of the relevant IR renormalon poles. Finally, in section~\ref{sect5},
we end with a summary and conclusions, and appendix~\ref{appA} provides the
extension of the work of ref.~\cite{jk86} to an arbitrary number of colours
$N_c$.

\section{Flavour non-diagonal vector and axial-vector correlators}\label{sect2}

We begin by investigating the dimension-6 OPE contributions to the two-point
correlation functions of flavour non-diagonal vector and axial-vector currents
$j_\mu^V(x)=(\bar u\gamma_\mu d)(x)$ and
$j_\mu^A(x)=(\bar u\gamma_\mu\gamma_5 d)(x)$ which are relevant for QCD
analyses of hadronic $\tau$ decays \cite{bnp92} and correspond to the charged
$\rho$ and $A_1$ mesons. For simplicity, massless light quarks will be assumed
in which case the correlators take the form
\begin{equation}
\Pi_{\mu\nu}^{V/A}(q) \,\equiv\, i\!\int\! {\rm d}x\, {\rm e}^{iqx}\,
\langle\Omega|T\{ j_\mu^{V/A}(x) j_\nu^{V/A}(0)^\dagger\}|\Omega\rangle
\,=\, \big( q_\mu q_\nu - g_{\mu\nu} q^2 \big)\, \Pi^{V/A}(q^2) \,.
\end{equation}
Here, $|\Omega\rangle$ denotes the full QCD vacuum and the second identity
follows because in the massless limit vector and axial-vector currents are
conserved.

In the framework of the OPE, the scalar functions $\Pi^{V/A}$ permit an
expansion in powers of $1/Q^2$ with $Q^2\equiv -q^2>0$ being in the Euclidean
region,
\begin{equation}
\label{ope}
\Pi^{V/A}(Q^2) \,=\, C_0(Q^2) + C_4(Q^2)\, \frac{\langle O_4 \rangle}{Q^4} +
C_6^{V/A}(Q^2)\, \frac{\langle O_6 \rangle}{Q^6} + \ldots \,.
\end{equation}
In writing eq.~\eqn{ope}, for simplicity, we have suppressed the vacuum state.
In the OPE, only the coefficient functions $C_i^{V/A}$ depend on the momentum,
while the operators $O_i$ are local. Both, coefficient functions and operators
depend, however, on the renormalisation scale $\mu$ which is not shown
explicitly. Furthermore, for flavour non-diagonal currents the purely
perturbative contribution $C_0(Q^2)$ is the same for vector and
axial vector.\footnote{For correlators of flavour-diagonal currents, which will
be discussed below, this is not the case.} In the massless case, this also
remains true for the dimension-4 contribution, which then only consists of the
gluon condensate $\langle G_{\mu\nu}^aG^{a\,\mu\nu}\rangle$.

Our main concern in this work will be the dimension-6 term which receives
contributions from the three-gluon condensate
$\langle g^3 f_{abc} G_{\mu\nu}^a G_{\;\;\;\lambda}^{b\,\nu} G^{c\,\lambda\mu}\rangle$
and four-quark condensates. As the three-gluon condensate does not arise at
leading order, below we concentrate only on the four-quark condensates. Their
contribution to $\Pi^{V/A}(Q^2)$ has been computed at the next-to-leading
order in refs.~\cite{ac94,lsc86}. For our following discussion, it will be
convenient to present the corresponding results for $V-A$ and $V+A$ correlation
functions, because in the former case the so-called penguin operator
contributions cancel. For $N_f=3$ light quark flavours and at $N_c=3$, one
then finds
\begin{equation}
\label{C6O6VmA}
C_6^{V-A}(Q^2)\,\langle O_6 \rangle \,=\,
4\pi^2 a_s \,\Big\{ \Big[\, 2 + \Big( \sfrac{25}{6} - L \Big) a_s \Big]
\langle Q_-^{\,o} \rangle -
\Big( \sfrac{11}{18} - \sfrac{2}{3} L \Big) a_s \langle Q_-^{\,s} \rangle
\Big\} \,,
\end{equation}
and
\begin{eqnarray}
C_6^{V+A}(Q^2)\,\langle O_6 \rangle \,&=&\,
-\,4\pi^2 a_s \,\Big\{ \Big[\, 2 + \Big( \sfrac{155}{24} - \sfrac{7}{2} L \Big)
a_s \Big] \langle Q_+^{\,o} \rangle +
\Big( \sfrac{11}{18} - \sfrac{2}{3} L \Big) a_s \langle Q_+^{\,s} \rangle +
\nn \\
\mvs
&& \hspace{17mm}
\Big[\, \sfrac{4}{9} + \Big( \sfrac{37}{36} - \sfrac{95}{162} L \Big) a_s \Big]
\langle Q_3 \rangle +
\Big( \sfrac{35}{108} - \sfrac{5}{18} L \Big) a_s \langle Q_4 \rangle + \nn \\
\mvs
&& \hspace{27mm}
\label{C6O6VpA}
\Big( \sfrac{14}{81} - \sfrac{4}{27} L \Big) a_s \langle Q_6 \rangle -
\Big( \sfrac{2}{81} + \sfrac{4}{27} L \Big) a_s \langle Q_7 \rangle \,.
\end{eqnarray}
Here, $a_s\equiv \as/\pi$, $L\equiv \ln Q^2/\mu^2$ and the constant terms of
order $a_s^2$ correspond to the choice of an anti-commuting $\gamma_5$ in $D$
space-time dimensions which can be made consistent as long as no traces with
an odd number of $\gamma_5$'s arise in the calculation \cite{ac94}.

The appearing four-quark operators are a subset which belong to the complete
basis that below will be required for their one-loop renormalisation:
\begin{eqnarray}
\label{Qbasis}
Q_V^{\,o} \,&=&\, (\bar u\gamma_\mu t^a d\bar d\gamma^\mu t^a u) \,, \quad
Q_A^{\,o} \,=\, (\bar u\gamma_\mu\gamma_5 t^a d\bar d\gamma^\mu\gamma_5 t^a u)
\label{QVQAo}
\,,\\
\tvs
Q_V^{\,s} \,&=&\, (\bar u\gamma_\mu d\bar d\gamma^\mu u) \,, \quad
Q_A^{\,s} \,=\, (\bar u\gamma_\mu\gamma_5 d\bar d\gamma^\mu\gamma_5 u) \,, \\
\tvs
Q_3 \,&\equiv&\, (\bar u\gamma_\mu t^a u + \bar d\gamma_\mu t^a d)
\!\sum\limits_{q=u,d,s} (\bar q\gamma^\mu t^a q) \,, \\
\tvs
Q_4 \,&\equiv&\, (\bar u\gamma_\mu\gamma_5 t^a u +
                  \bar d\gamma_\mu\gamma_5 t^a d)
\!\sum\limits_{q=u,d,s} (\bar q\gamma^\mu\gamma_5 t^a q) \,, \\
\tvs
Q_5 \,&\equiv&\, (\bar u\gamma_\mu u + \bar d\gamma_\mu d)
\!\sum\limits_{q=u,d,s} (\bar q\gamma^\mu q) \,, 
\end{eqnarray}
\begin{eqnarray}
Q_6 \,&\equiv&\, (\bar u\gamma_\mu\gamma_5 u + \bar d\gamma_\mu\gamma_5 d)
\!\sum\limits_{q=u,d,s} (\bar q\gamma^\mu\gamma_5 q) \,, \\
\tvs
Q_7 \,&\equiv&\, \sum\limits_{q=u,d,s} (\bar q\gamma_\mu t^a q)
\!\sum\limits_{q^\prime=u,d,s} (\bar q^\prime\gamma^\mu t^a q^\prime) \,, \\
\tvs
Q_8 \,&\equiv&\, \sum\limits_{q=u,d,s} (\bar q\gamma_\mu\gamma_5 t^a q)
\!\sum\limits_{q^\prime=u,d,s} (\bar q^\prime\gamma^\mu\gamma_5 t^a q^\prime)
\,, \\
\tvs
Q_9 \,&\equiv&\, \sum\limits_{q=u,d,s} (\bar q\gamma_\mu q)
\!\sum\limits_{q^\prime=u,d,s} (\bar q^\prime\gamma^\mu q^\prime) \,, \\
\tvs
Q_{10} \,&\equiv&\, \sum\limits_{q=u,d,s} (\bar q\gamma_\mu\gamma_5 q)
\!\sum\limits_{q^\prime=u,d,s} (\bar q^\prime\gamma^\mu\gamma_5
q^\prime) \,. \label{Q10}
\end{eqnarray}
The operators $Q_{V/A}^{\,o}$ and $Q_{V/A}^{\,s}$ are usually termed
current-current operators and $Q_3$ to $Q_{10}$ penguin operators. In addition, we define the four current-current operators which appear in eqs.~\eqn{C6O6VmA}
and \eqn{C6O6VpA}.
\begin{equation}
\label{Qpmos}
Q_\pm^{\,o} \,\equiv\, Q_V^{\,o} \pm Q_A^{\,o} \,, \qquad
Q_\pm^{\,s} \,\equiv\, Q_V^{\,s} \pm Q_A^{\,s} \,. 
\end{equation}

Next, we investigate the scale dependence of a general term $R_O$ in the OPE,
corresponding to a set of operators $\vec O$ with equal dimension,
\begin{equation}
\label{Rope}
R_O \,=\, \vec C^{\,T\!}(\mu)\,\langle \vec O(\mu)\rangle \,,
\end{equation}
where now the renormalisation scale $\mu$ is displayed explicitly and the
potential dependence on other dimensionful parameters is implicit.
For vector and axial-vector currents, the renormalisation scale dependence of
the correlator only arises from the purely perturbative contribution. Hence,
$R_O$ should not depend on $\mu$, and it immediately follows that
\begin{equation}
\label{Rrge}
\biggl[ \mu\,\frac{{\rm d}}{{\rm d}\mu}\,\vec C^{\,T\!}(\mu) \biggr]
\langle \vec O(\mu)\rangle \,=\, -\,\vec C^{\,T\!}(\mu)\Biggl[
\mu\,\frac{{\rm d}}{{\rm d}\mu}\,\langle \vec O(\mu)\rangle \Biggr] \,.
\end{equation}
The anomalous dimension matrix $\hat\gamma_O$ of the operator matrix
elements can be defined by
\begin{equation}
\label{gammaO}
-\,\mu\,\frac{{\rm d}}{{\rm d}\mu}\,\langle \vec O(\mu)\rangle \,\equiv\,
\hat\gamma_O(a_\mu)\,\langle \vec O(\mu)\rangle \,,
\end{equation}
with $a_\mu\equiv a_s(\mu)$.
If the bare and renormalised operator matrix elements are related by
\begin{equation}
\label{Obare}
\langle\vec O^B\rangle \,\equiv\, \hat Z_O(\mu)\,\langle\vec O(\mu)\rangle \,,
\end{equation}
it follows that the anomalous dimension matrix can be computed from the
renormalisation matrix $\hat Z_O(\mu)$ via
\begin{equation}
\label{gammaOZ}
\hat\gamma_O(a_\mu) \,=\, \hat Z_O^{-1}(\mu)\,\mu\,\frac{{\rm d}}{{\rm d}\mu}\,
\hat Z_O(\mu) \,.
\end{equation}
Plugging eq.~\eqn{gammaO} into the RGE for $R$, eq.~\eqn{Rrge}, one obtains
an RGE that has to be satisfied by the coefficient functions $\vec C(\mu)$,
\begin{equation}
\label{Crge}
\mu\,\frac{{\rm d}}{{\rm d}\mu}\,\vec C(\mu) \,=\, \hat\gamma_O^T(a_\mu)\,\vec C(\mu) \,.
\end{equation}
This equation shall be checked for the coefficient functions of the dimension-6 operators in eqs.~\eqn{C6O6VmA} and \eqn{C6O6VpA}.
Furthermore, below it will be convenient to consider the anomalous
dimension matrix in a linearly transformed basis. If the transformed basis
$\langle\vec O^\prime(\mu)\rangle$ of operator matrix elements is defined by
\begin{equation}
\label{Qvec}
\langle\vec O^\prime(\mu)\rangle \,\equiv\,
\hat T\,\langle\vec O(\mu)\rangle \,,
\end{equation}
the corresponding transformed anomalous dimension matrix takes the form
\begin{equation}
\label{gammaQ}
\hat\gamma_{O^\prime} \,=\, \hat T\,\hat\gamma_O\,\hat T^{-1} \,.
\end{equation}

\begin{figure}[thb]
\begin{center}
\includegraphics[height=3.2cm]{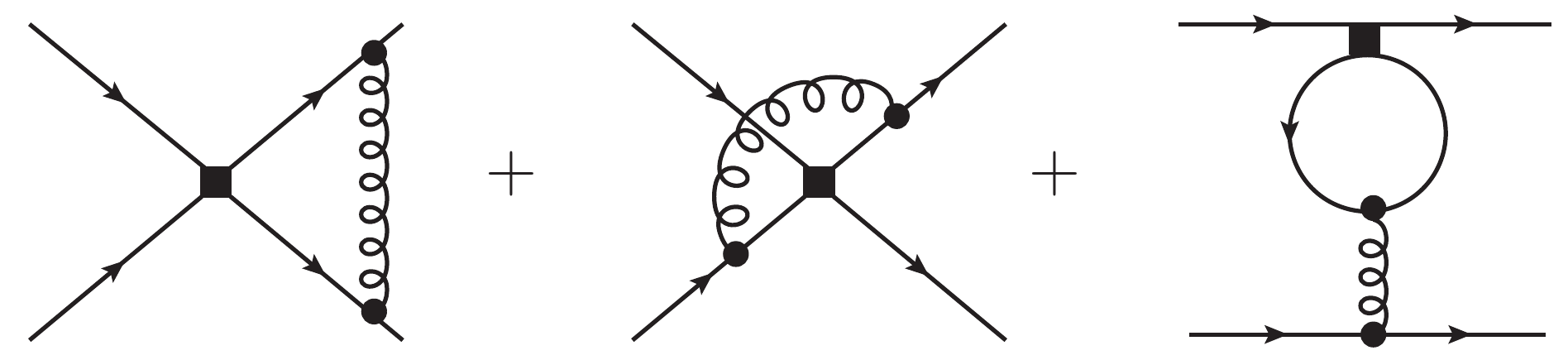}
\caption{Exemplary one-loop current-current and penguin diagrams that have to
be calculated in the process of obtaining the leading-order anomalous dimension
matrix of four-quark operators. Quark self-energy diagrams are not displayed.
\label{fig1}}
\end{center}
\end{figure}

The calculation of one-loop anomalous dimension matrices for dimension-6
four-quark operators is fairly standard and details can for example be found
in ref.~\cite{mm83}. We have performed the actual computation in two ways:
firstly, by explicit calculation of the diagrams of figure~\ref{fig1}, and
secondly, by relating the appearing operators to the complete basis of
dimension-6 four-quark operators without derivatives in the case of three
quark flavours that had been employed in ref.~\cite{jk86}. Further details
on the second approach can be found in appendix~\ref{appA}. The computation
has been performed in a general covariant gauge in order to explicitly verify
gauge invariance.

Expanding the anomalous dimension matrix in a power series in $a_s$,
\begin{equation}
\hat\gamma_O(a_s) \,=\, a_s\,\hat\gamma_O^{(1)} +
a_s^2\,\hat\gamma_O^{(2)} + \ldots \,,
\end{equation}
the leading-order anomalous dimension matrix corresponding to the operators
$Q_-\equiv (Q_-^o,\, Q_-^s)$ appearing in eq.~\eqn{C6O6VmA} is found to be
\begin{equation}
\label{gammaQm}
\hat\gamma_{Q_-}^{(1)} \,=\,
\left(\! \begin{array}{cc}
-\,\sfrac{3N_c}{2}+\sfrac{3}{N_c} & -\,\sfrac{3C_F}{2N_c} \\[2mm]
-\,3 & 0 \end{array} \right) .
\end{equation}
At this order, the set of two operators is closed under renormalisation,
meaning that no additional operators are generated through the diagrams that
have to be computed.\footnote{This is only true when working with a strictly
anti-commuting $\gamma_5$, and projecting out evanescent operators, which,
however, is admissible at the leading order. Further discussion can be found
in ref.~\cite{ac94}.} Employing $\hat\gamma_{Q_-}^{(1)}$ and the coefficient
function $C_6^{V-A}(Q^2)$ of eq.~\eqn{C6O6VmA}, it is a simple matter to
confirm that the RGE \eqn{Crge} is indeed satisfied.

Likewise, the anomalous dimension matrix for the operators appearing in the
$V+A$ case of eq.~\eqn{C6O6VpA} can be calculated. Here, three additional
operators have to be added in the course of renormalisation, and at the leading
order a closed set can be chosen as
$Q_+\equiv(Q_+^o,\, Q_+^s,\, Q_3,\, Q_4,\, Q_6,\, Q_7,\, Q_8,\, Q_9,\, Q_{10})$.
In general, also the operator $Q_5$ of the basis presented above arises.
However, in four dimensions, one operator in the full set is redundant and can
be expressed through the others by means of Fierz transformations. Since $Q_5$
does not appear in the OPE expression \eqn{C6O6VpA}, we have rewritten it
through the remaining operators (see eq.~(\ref{OperatorRelation})). The
anomalous dimension matrix then takes the form

\begin{displaymath}
\hspace{-30mm} \hat\gamma_{Q_+}^{(1)} \,=\,
\left(\! \begin{array}{cccc}
-\,\sfrac{3}{N_c} & \sfrac{3C_F}{2N_c} & -\,\sfrac{1}{3N_c} & 0 \\[2mm]
3 & 0 & \sfrac{2}{3} & 0 \\[2mm]
0 & 0 & \sfrac{N_f}{3}-\sfrac{3N_c}{4}-\sfrac{1}{3N_c} & \sfrac{3N_c}{4}-\sfrac{3}{N_c} \\[2mm]
\sfrac{3}{2}+\sfrac{3}{2N_c}\; & \;-\,\sfrac{3C_F}{2N_c}\; & \sfrac{3N_c}{4}+\sfrac{3}{2}-\sfrac{11}{6N_c} & \;-\,\sfrac{3N_c}{4}+\sfrac{3}{2}+\sfrac{3}{2Nc} \\[2mm]
0 & 0 & \sfrac{11}{3} & 0 \\[2mm]
0 & 0 & 0 & 0 \\[2mm]
0 & 0 & 0 & 0 \\[2mm]
0 & 0 & 0 & 0 \\[2mm]
0 & 0 & 0 & 0 \end{array} \right.
\end{displaymath}

\vspace{5mm}
\begin{equation}
\label{gammaQp}
\hspace{30mm} \left. \begin{array}{ccccc}
0 & 0 & 0 & 0 & 0 \\[2mm]
0 & 0 & 0 & 0 & 0 \\[2mm]
\sfrac{3C_F}{2N_c} & \sfrac{2}{3} & 0 & 0 & 0 \\[2mm]
-\,\sfrac{3C_F}{2N_c} & -\,\sfrac{3}{4}-\sfrac{3}{4N_c} & -\,\sfrac{3}{4}-\sfrac{3}{4N_c}& \;\sfrac{3C_F}{4N_c}\; & \;\sfrac{3C_F}{4N_c}\; \\[2mm]
0 & 0 & 0 & 0 & 0 \\[2mm]
0 & \sfrac{2N_f}{3}-\sfrac{3N_c}{4}-\sfrac{1}{3N_c}\; & \;\sfrac{3N_c}{4}-\sfrac{3}{N_c} & 0 & \sfrac{3C_F}{2N_c} \\[2mm]
0 & \sfrac{3N_c}{4}-\sfrac{10}{3N_c} & -\,\sfrac{3N_c}{4} & \sfrac{3C_F}{2N_c} & 0 \\[2mm]
0 & \sfrac{2}{3} & 3 & 0 & 0 \\[2mm]
0 & \sfrac{11}{3} & 0 & 0 & 0 \end{array} \right) .
\end{equation}
In ref.~\cite{bbk97}, and section~3.2.3 of ref.~\cite{ben98}, the SU(3)
flavour-singlet operators $Q_7$ to $Q_{10}$ arose in an investigation of the
structure of the leading UV renormalon at $u=-1$ for the vector correlator.
Though not directly related to our study, the relevant leading-order anomalous
dimensions correspond to the $4\!\times\!4$ sub-matrix for $Q_7$ to $Q_{10}$,
and comparing to the results of refs.~\cite{bbk97,ben98}, we find complete
agreement.\footnote{The same subset of anomalous dimensions was already
considered in a previous attempt to investigate the $u=3$ renormalon structure
\cite{dm10}.} Regarding the RGE, again, it is straightforward to verify that
by using $\hat\gamma_{Q_+}^{(1)}$ and the coefficient function of
eq.~\eqn{C6O6VpA}, the RGE \eqn{Crge} is satisfied to leading order.

For the subsequent discussion, it will be convenient to consider a basis of
four-quark operators in which the leading-order anomalous dimension matrix is
diagonal.  From linear algebra it is well known that the anomalous dimension
matrix $\hat\gamma_O^{(1)}$ can be diagonalised by a matrix $\hat V$, which as
columns contains the eigenvectors of $\hat\gamma_O^{(1)}$, in the following
fashion:
\begin{equation}
\hat\gamma_D^{(1)} \,=\, \hat V^{-1}\,\hat\gamma_O^{(1)}\,\hat V \,. 
\label{Diag}
\end{equation}
The diagonal entries of $\hat\gamma_D^{(1)}$ then correspond to the
eigenvalues of $\hat\gamma_O^{(1)}$. Furthermore, the operator basis with
$\hat\gamma_D^{(1)}$ as the leading-order anomalous dimension matrix is given
by $\hat V^{-1}\vec O$. Rewriting the term $R_O$ of \eqn{Rope} in the OPE,
\begin{equation}
R_O \,=\, \vec C^{\,T\!}(\mu)\,\hat V\;\hat V^{-1}
\langle \vec O(\mu)\rangle \,,
\end{equation}
the logarithms in $\vec C(\mu)$ can be resummed to leading order by solving
the RGE \eqn{Crge}, leading to
\begin{equation}
\label{ROres}
R_O \,=\, \vec C^{\,T\!}(Q)\,\hat V\biggl[\Big(
\frac{a_Q}{a_\mu}\Big)^{\!\vec\gamma_D^{(1)}/\beta_1}\biggr]_D\,\hat V^{-1}
\langle \vec O(\mu)\rangle \,.
\end{equation}
The somewhat condensed notation in \eqn{ROres} should be read as follows:
$\vec\gamma_D^{(1)}$ is a vector containing the eigenvalues of
$\hat\gamma_O^{(1)}$ ordered according to the eigenvectors in $\hat V$. Then
$[..]_D$ is the diagonal matrix which contains as diagonal entries the
ratios of $a_s$ to the power of elements in $\vec\gamma_D^{(1)}/\beta_1$.
Finally, $\beta_1=11N_c/6-N_f/3$ is the leading coefficient in the expansion
of the QCD $\beta$-function
\begin{equation}
\label{betafun}
-\mu\,\frac{{\rm d}a_\mu}{{\rm d}\mu} \,\equiv\, \beta(a_\mu) \,=\,
\beta_1 a_\mu^2 + \beta_2\, a_\mu^3 + \ldots \,.
\end{equation}
The generalisation of \eqn{ROres} to next-to-leading order is slightly
non-trivial because anomalous dimension matrices at different couplings do
not commute, but it can for example be found in refs.~\cite{bur80,bjlw92}.

Numerically, at $N_c=N_f=3$ the eigenvalues of the anomalous dimension matrices
$\hat\gamma_{Q_-}^{(1)}$ and $\hat\gamma_{Q_+}^{(1)}$, ordered in increasing
value, are found to be
\begin{eqnarray}
\label{evQmSU3}
\vec\gamma_{D,\,Q_-}^{(1)} \,&=&\, (-4,\, 0.5) \,, \\
\tvs
\label{evQpSU3}
\vec\gamma_{D,\,Q_+}^{(1)} \,&=&\, (-3.611,\, -3.387,\, -1.878,\, -1.494,\,
0.538,\, 0.567,\, 1,\, 1.340,\, 1.703) \,.
\end{eqnarray}
Besides the eigenvalue $1$, the entries in $\vec\gamma_{D,\,Q_+}^{(1)}$
are found as the roots of the two quartic polynomials
$176 - 316 z - 101 z^2 + 130 z^3 + 36 z^4$ and
$88 - 122 z -  91 z^2 +  47 z^3 + 18 z^4$.
The corresponding eigenvectors have been collected in appendix~\ref{appB}.
Regarding the operator combinations $\hat V_{Q_+}^{-1}\vec Q_+$, it is found
that four of them just include the operators $Q_7$ to $Q_{10}$, and a further
combination misses the operator $Q_6$, while the remaining ones contain all
contributing operators. The eigenvalues corresponding to the combinations only
containing $Q_7$ to $Q_{10}$ (entries 2, 4, 6 and 9) agree with the result of
table~1 in references~\cite{ben98,bbk97} where the same operators were
considered to study the structure of the leading UV renormalon.

To conclude this section, we investigate the case of flavour SU(2).
Then, the closed $(V+A)$ operator basis is given by
$\ol Q_+\equiv(Q_+^o,\, Q_+^s,\, \ol Q_3,\, \ol Q_4,\, \ol Q_6)$, where in
the penguin operators the strange quark is removed and hence the operators
$\ol Q_3$ to $\ol Q_6$ are analogous to $Q_7$ to $Q_{10}$ with the sums just
running over up and down quarks. The corresponding anomalous dimension matrix
is found to be
\begin{equation}
\label{gammaQpSU2}
\hat\gamma_{\ol Q_+}^{(1)} \,=\,
\left(\! \begin{array}{ccccc}
-\,\sfrac{3}{N_c} & \sfrac{3C_F}{2N_c} & -\,\sfrac{1}{3N_c} & 0 & 0 \\[2mm]
3 & 0 & \sfrac{2}{3} & 0 & 0 \\[2mm]
0 & 0 & \sfrac{2N_f}{3}-\sfrac{3N_c}{4}-\sfrac{1}{3N_c} & \sfrac{3N_c}{4}-\sfrac{3}{N_c} & \sfrac{3C_F}{2N_c} \\[2mm]
3+\sfrac{3}{N_c}\; & \;-\,\sfrac{3C_F}{N_c}\; & \sfrac{3N_c}{4}+\sfrac{3}{2}-\sfrac{11}{6N_c} & \;-\,\sfrac{3N_c}{4}+\sfrac{3}{2}+\sfrac{3}{2Nc}\; & \;-\,\sfrac{3C_F}{2N_c} \\[2mm]
0 & 0 & \sfrac{11}{3} & 0 & 0 \end{array} \right) ,
\end{equation}
with the eigenvalues
\begin{equation}
\vec\gamma_{D,\,\ol Q_+}^{(1)} \,=\, (-3.521,\, -1.751,\, 0.549,\,
1,\, 1.445)
\end{equation}
at $N_c=3$ and $N_f=2$. It is observed that the eigenvalues are slightly
shifted with respect to the SU(3) case \eqn{evQpSU3}, but span approximately
the same range.

\section{Flavour diagonal vector and axial-vector correlators}\label{sect3}

In this section, we now move to a discussion of dimension-6 OPE contributions
to the flavour-diagonal vector and axial-vector currents. For definiteness, we
consider the up-quark case for which the corresponding currents are given by
$j_\mu^V(x)=(\bar u\gamma_\mu u)(x)$ and
$j_\mu^A(x)=(\bar u\gamma_\mu\gamma_5 u)(x)$. The vector current for example
is relevant in $e^+e^-$ scattering to hadrons and carries the quantum numbers
of the neutral $\rho$ meson.

To leading order, the contribution $C_6^{V\pm A}(Q^2)\langle O_6\rangle$ to
the correlation function are the same as eqs.~\eqn{C6O6VmA} and \eqn{C6O6VpA}
with the current-current operators now being given by
\begin{eqnarray}
\label{Qccsing}
Q_V^{\,o} \,&=&\, (\bar u\gamma_\mu t^a u\bar u\gamma^\mu t^a u) \,, \quad
Q_A^{\,o} \,=\, (\bar u\gamma_\mu\gamma_5 t^a u\bar u\gamma^\mu\gamma_5 t^a u)
\,,\\
\tvs
Q_V^{\,s} \,&=&\, (\bar u\gamma_\mu u\bar u\gamma^\mu u) \,, \quad
Q_A^{\,s} \,=\, (\bar u\gamma_\mu\gamma_5 u\bar u\gamma^\mu\gamma_5 u) \,.
\end{eqnarray}
Unfortunately, to our knowledge, for the flavour-diagonal correlators, the
next-to-leading order corrections to the dimension-6 four-quark operators are
not available. The next important difference to the non-diagonal case is the
fact that the two current-current operators $Q_-^{o,s}$ are not anymore closed
under renormalisation, but all ten operators including the eight penguin
operators are required. Also, now all operators are linearly independent.
Hence, the complete set reads
$Q_-^{\rm diag}\equiv(Q_-^o,\, Q_-^s,\, Q_3,\, Q_4,\, Q_5,\, Q_6,\, Q_7,\, Q_8,\, Q_9,\, Q_{10})$.
For this set of operators, the leading-order anomalous dimension matrix is
found to be

\begin{displaymath}
\hspace{-30mm} \hat\gamma_{Q_-^{\rm diag}}^{(1)} \,=\,
\left(\! \begin{array}{ccccc}
-\,\sfrac{3N_c}{2}+\sfrac{3}{N_c}\; & \;-\,\sfrac{3C_F}{2N_c} & \sfrac{2}{3} & 0 & 0 \\[2mm]
-\,3 & 0 & 0 & 0 & 0 \\[2mm]
0 & 0 & \sfrac{N_f}{3}-\sfrac{3N_c}{4}-\sfrac{1}{3N_c}\; & \;\sfrac{3N_c}{4}-\sfrac{3}{N_c} & 0 \\[2mm]
0 & 0 & \sfrac{3N_c}{4}-\sfrac{10}{3N_c} & -\,\sfrac{3N_c}{4} & \sfrac{3C_F}{2N_c} \\[2mm]
0 & 0 & \sfrac{2}{3} & 3 & 0 \\[2mm]
0 & 0 & \sfrac{11}{3} & 0 & 0 \\[2mm]
0 & 0 & 0 & 0 & 0 \\[2mm]
0 & 0 & 0 & 0 & 0 \\[2mm]
0 & 0 & 0 & 0 & 0 \\[2mm]
0 & 0 & 0 & 0 & 0 \end{array} \right.
\end{displaymath}

\vspace{5mm}
\begin{equation}
\label{gammaQmsing}
\hspace{30mm} \left. \begin{array}{ccccc}
0 & 0 & 0 & 0 & 0 \\[2mm]
0 & 0 & 0 & 0 & 0 \\[2mm]
\sfrac{3C_F}{2N_c} & \sfrac{1}{3} & 0 & 0 & 0 \\[2mm]
0 & 0 & 0 & 0 & 0 \\[2mm]
0 & 0 & 0 & 0 & 0 \\[2mm]
0 & 0 & 0 & 0 & 0 \\[2mm]
0 & \sfrac{2N_f}{3}-\sfrac{3N_c}{4}-\sfrac{1}{3N_c}\; & \;\sfrac{3N_c}{4}-\sfrac{3}{N_c} & 0 & \sfrac{3C_F}{2N_c} \\[2mm]
0 & \sfrac{3N_c}{4}-\sfrac{10}{3N_c} & -\,\sfrac{3N_c}{4} & \sfrac{3C_F}{2N_c} & 0 \\[2mm]
0 & \sfrac{2}{3} & 3 & 0 & 0 \\[2mm]
0 & \sfrac{11}{3} & 0 & 0 & 0 \end{array} \right) .
\end{equation}

For the flavour-diagonal $V+A$ correlator, the two current-current operators
$Q_+^o$ and $Q_+^s$ are directly linearly related via Fierz transformations,
with the relation being given by
\begin{equation}
Q_+^o \,=\, \frac{1}{2}\,\Big( 1 - \frac{1}{N_c} \Big)\, Q_+^s \,.
\end{equation}
Choosing to remove the $Q_+^s$ operator, like in the non-diagonal case, we are
again left we a set of nine operators, consisting of
$Q_+^{\rm diag}\equiv(Q_+^o,\, Q_3,\, Q_4,\, Q_5,\, Q_6,\, Q_7,\, Q_8,\, Q_9,\, Q_{10})$.
The corresponding leading-order anomalous dimension matrix is then found to be

\begin{displaymath}
\hspace{-30mm} \hat\gamma_{Q_+^{\rm diag}}^{(1)} \,=\,
\left(\! \begin{array}{cccc}
\sfrac{3}{2}-\sfrac{3}{2N_c} & \sfrac{2}{3}-\sfrac{2}{3N_c} & 0 & 0 \\[2mm]
0 & \sfrac{N_f}{3}-\sfrac{3N_c}{4}-\sfrac{1}{3N_c}\; & \;\sfrac{3N_c}{4}-\sfrac{3}{N_c} & 0 \\[2mm]
0 & \sfrac{3N_c}{4}-\sfrac{10}{3N_c} & -\,\sfrac{3N_c}{4} & \sfrac{3C_F}{2N_c} \\[2mm]
0 & \sfrac{2}{3} & 3 & 0 \\[2mm]
0 & \sfrac{11}{3} & 0 & 0 \\[2mm]
0 & 0 & 0 & 0 \\[2mm]
0 & 0 & 0 & 0 \\[2mm]
0 & 0 & 0 & 0 \\[2mm]
0 & 0 & 0 & 0 \end{array} \right.
\end{displaymath}

\vspace{5mm}
\begin{equation}
\label{gammaQpsing}
\hspace{30mm} \left. \begin{array}{ccccc}
0 & 0 & 0 & 0 & 0 \\[2mm]
\sfrac{3C_F}{2N_c} & \sfrac{1}{3} & 0 & 0 & 0 \\[2mm]
0 & 0 & 0 & 0 & 0 \\[2mm]
0 & 0 & 0 & 0 & 0 \\[2mm]
0 & 0 & 0 & 0 & 0 \\[2mm]
0 & \sfrac{2N_f}{3}-\sfrac{3N_c}{4}-\sfrac{1}{3N_c}\; & \;\sfrac{3N_c}{4}-\sfrac{3}{N_c} & 0 & \sfrac{3C_F}{2N_c} \\[2mm]
0 & \sfrac{3N_c}{4}-\sfrac{10}{3N_c} & -\,\sfrac{3N_c}{4} & \sfrac{3C_F}{2N_c} & 0 \\[2mm]
0 & \sfrac{2}{3} & 3 & 0 & 0 \\[2mm]
0 & \sfrac{11}{3} & 0 & 0 & 0 \end{array} \right) .
\end{equation}

Let us again investigate the eigenvalues for the anomalous dimension matrices
$\hat\gamma_{Q_-^{\rm diag}}^{(1)}$ and $\hat\gamma_{Q_+^{\rm diag}}^{(1)}$.
Numerically, at $N_c=N_f=3$ and ordered in increasing value, they are found to
be
\begin{eqnarray}
\label{evQmsing}
\vec\gamma_{D,\,Q_-^{\rm diag}}^{(1)} \,&=&\, (-4,\, -3.611,\, -3.387,\,
-1.878,\, -1.494,\, 0.5,\, 0.538,\, 0.567,\, 1.340, 1.703) , \\
\tvs
\label{evQpsing}
\vec\gamma_{D,\,Q_+^{\rm diag}}^{(1)} \,&=&\, (-3.611,\, -3.387,\, -1.878,\,
-1.494,\, 0.538,\, 0.567,\, 1,\, 1.340,\, 1.703) \,.
\end{eqnarray}
We first note, that the eigenvalues corresponding to
$\hat\gamma_{Q_+^{\rm diag}}^{(1)}$ are identical to the ones for
$\hat\gamma_{Q_+}^{(1)}$ of eq.~\eqn{evQpSU3}. This will have implications for
the renormalon structure to be discussed in the next section. Furthermore, as
can be inferred from the eigenvectors of eq.~\eqn{evVpA}, the eigenvalue ``1''
corresponds to the current-current operators while the remaining ones involve
penguin operators. The eigenvalues for $\hat\gamma_{Q_-^{\rm diag}}^{(1)}$,
on the other hand, include the ones from the non-diagonal current-current
operators given in eq.~\eqn{evQmSU3} as well as again the ones involving the
penguins.

\section{Renormalon structure of dimension-6 four-quark operators}\label{sect4}

In this section, the perturbative ambiguities that are connected to the
dimension-6 four-quark OPE contributions, shall be investigated. The
discussion closely follows section~3.3 of ref.~\cite{ben98} and section~5
of ref.~\cite{bj08}. 

Before investigating the renormalon structure, however, we have to briefly
review the corresponding nomenclature. To begin, we introduce the Adler
functions for non-diagonal vector and axial-vector correlators that are physical
quantities in the sense that they are independent of renormalisation scale and
scheme:
\begin{equation}
D^{V/A}(Q^2) \,\equiv\, -\,Q^2\,\frac{{\rm d}}{{\rm d}Q^2}\,\Pi^{V/A}(Q^2) \,.
\end{equation}
Next, we define the purely perturbative function $\wh D_0(Q^2)$ through the
relation
\begin{equation}
D_0^{V/A}(Q^2) \,\equiv\, D_0(Q^2) \,\equiv\,
{\cal N}\,\big[\, 1 + \wh D_0(Q^2) \,\big] \,,
\end{equation}
such that $\wh D_0(Q^2)$ starts at order $\as$, and ${\cal N}=N_c/(12\pi^2)$ is
the common global normalisation. Because the perturbative part is identical for
non-diagonal vector and axial-vector correlators, $\wh D_0(Q^2)$ is the same in
both cases.

Assuming a positive coupling $a_Q$, the Borel transform $B[\wh D_0](u)$ of
$\wh D_0(a_Q)$ is defined by 
\begin{equation}
\wh D_0(a_Q) \,\equiv\, \frac{2\pi}{\beta_1}\! \int\limits_0^\infty\!
{\rm d}u\, {\rm e}^{-\frac{2u}{\beta_1 a_Q}} B[\wh D_0](u) \,.
\end{equation}
Because $D_0$ is dimensionless, its $Q^2$ dependence only arises via $a_Q$.
Taylor expanding $B[\wh D_0](u)$ and integrating term by term reproduces the
perturbation series expansion of $\wh D_0(a_Q)$. The integral on the right-hand
side is, however, only well defined if $B[\wh D_0](u)$ has no poles or cuts on
the positive real $u$ axis which is not the case for the QCD Adler function.
If poles and/or cuts, also termed infrared (IR) renormalon poles, are present
on the integration interval, ambiguities appear since on  has to  specify how
the poles are treated. (For example with the principal-value prescription.)
It is generally assumed that those ambiguities should cancel against
corresponding ambiguities in the OPE terms, in order to ensure that
$D^{V/A}(Q^2)$ is physical.

From the generic structure of a term in the OPE, the general form of an
infrared renormalon pole corresponding to an operator $O_d$ of dimension $d$
can be deduced, with the finding \cite{ben98,bj08}
\begin{equation}
\label{IRpole}
B[D_0^{\rm IR}](u) \,=\, \frac{d_p^{\rm IR}}{(p-u)^{\kappa}}\,
\big[\, 1 + {\cal O}(p-u) \,\big] \,,
\end{equation}
where $d_p^{\rm IR}$ are non-perturbative normalisations (residua) of the
renormalon poles that cannot be determined from renormalisation group
arguments. On the other hand,
\begin{equation}
p \,=\, \frac{d}{2} \,, \quad
\kappa \,=\, 1 - \delta + 2p\,\frac{\beta_2}{\beta_1^2} -
\frac{\gamma_{O_d}^{(1)}}{\beta_1} \,.\label{PoleStrength}
\end{equation}
Here, $\delta$ is the leading power in $a_s$ of the coefficient function,
$\delta=1$ in our case of eqs.~\eqn{C6O6VmA} and \eqn{C6O6VpA}, and
$\gamma_{O_d}^{(1)}$ is the leading order anomalous dimension of the
operator $O_d$. Hence, the strength of the pole $\kappa$ (as well as
all sub-leading terms) only depends on coefficients of the RG functions
(and coefficient functions of the OPE contributions).

Before further discussing our particular case of the renormalon structure
of dimension-6 four-quark operators, we have to investigate which operators
contribute to the perturbative ambiguity. To this end, we rewrite the
current-current operators of eqs.~\eqn{Qpmos} by means of Fierz transformations
and separating the quark chiralities, which in the case of $Q_-^{o,s}$ leads to
\begin{eqnarray}
\label{QmoLR}
Q_-^o \,&=&\, -\,\frac{4C_F}{N_c} \big( \bar u_L u_R \bar d_R d_L +
(L \leftrightarrow R) \big) +
\frac{4}{N_c} \big( \bar u_L t^a u_R \bar d_R t^a d_L +
(L \leftrightarrow R) \big) , \\
\tvs
\label{QmsLR}
Q_-^s \,&=&\, -\,\frac{4}{N_c} \big( \bar u_L u_R \bar d_R d_L +
(L \leftrightarrow R) \big) -
8 \big( \bar u_L t^a u_R \bar d_R t^a d_L +
(L \leftrightarrow R) \big) .
\end{eqnarray}
From \eqn{QmoLR} and \eqn{QmsLR} it is seen that the operators $Q_-^{o,s}$ are
order parameters of the SU$(N_f)_A$ symmetry breaking and hence they cannot
contribute to the perturbative ambiguity. This is also obvious from the fact
that for non-diagonal quark currents in $V-A$, the purely perturbative
contribution cancels and hence no related ambiguity can arise. This is
different for the operators appearing in $V+A$, where rewriting the
current-current operators $Q_+^{o,s}$ results in
\begin{eqnarray}
\label{QpoLR}
Q_+^o \,&=&\, \frac{2C_F}{N_c} \big(
\bar u_L \gamma_\mu u_L \bar d_L \gamma^\mu d_L + (L \rightarrow R) \big) -
\frac{2}{N_c} \big( \bar u_L \gamma_\mu t^a u_L \bar d_L \gamma^\mu t^a d_L +
(L \rightarrow R) \big) , \\
\tvs
\label{QpsLR}
Q_+^s \,&=&\, \frac{2}{N_c} \big(
\bar u_L \gamma_\mu u_L \bar d_L \gamma^\mu d_L + (L \rightarrow R) \big) +
4 \big( \bar u_L \gamma_\mu t^a u_L \bar d_L \gamma^\mu t^a d_L +
(L \rightarrow R) \big) .
\end{eqnarray}
Those two operators can and do have a perturbative ambiguity which is reflected
in the perturbative series  of the unit operator of the OPE for the $V+A$
correlator. The same holds true also for the penguin operators $Q_3$ to
$Q_{10}$.

Regarding the pole structure of IR renormalon poles corresponding to
dimension-6 operators in the OPE, the poles are located at $p=u=3$. Furthermore,
at $N_c=3$ and for three quark flavours, the exponent $\kappa$ takes the value
\begin{equation}
\kappa \,=\, \frac{64}{27} - \frac{2}{9}\,\gamma_{O_6}^{(1)} \,.
\end{equation}
The strongest singularity is assumed for the most negative eigenvalue in
$\vec\gamma_{D,\,Q_+}^{(1)}$ of eq.~\eqn{evQpSU3}, leading to
$\kappa=3.173$, while the weakest pole corresponds to  an exponent
$\kappa=1.992$.

Let us briefly compare these findings with the large-$\beta_0$
approximation.\footnote{For historical reasons, we speak about the
``large-$\beta_0$'' approximation, although in the notation employed in this
work, the leading coefficient of the $\beta$-function is termed $\beta_1$.}
This approximation can be obtained by considering the large-$N_f$ limit and
then replacing $-N_f/3 \rightarrow \beta_1$. In the anomalous dimension matrix
of eq.~\eqn{gammaQp} then only two entries, $-\beta_1$ and $-2\beta_1$, for the
operators $Q_3$ and $Q_7$, respectively, are left on the diagonal. Furthermore,
the $\beta_2$ term in the exponent $\kappa$ is absent, so that $\kappa=1$ or
$2$. This is precisely in line with the result of the large-$\beta_0$
approximation for the Adler function that at $u=3$ a linear and  a quadratic pole
is induced by dimension-6 operators \cite{ben98}. Hence, it is observed that
full QCD tends to make the $u=3$ IR renormalon poles stronger in comparison
to the large-$N_f$ limit. This conclusion also remains true for the operators
appearing in the case of flavour-diagonal correlators, because the eigenvalues
of the anomalous dimension matrices are identical to the ones of the flavour
non-diagonal case.

\section{Conclusions}\label{sect5}

In this work, we revisited the calculation of one-loop anomalous dimension
matrices of four-quark operators that enter vector and axial-vector correlators
in QCD. We studied flavour diagonal and non-diagonal currents. Our results are
given for an arbitrary numbers of quark colours, $N_c$, and flavours, $N_f$.

Explicit results for the anomalous dimension matrices of $V-A$ and $V+A$
flavour non-diagonal operators are given in eqs.~\eqn{gammaQm} and \eqn{gammaQp}.
In the $V-A$ case, the set of two operators of eq.~\eqn{Qpmos} is closed under
renormalisation. The $V+A$ case requires the inclusion of 7 additional
penguin-type operators to obtain a minimal closed set. We also presented
the SU(2) flavour non-diagonal $V+A$ anomalous dimension matrix in
eq.~\eqn{gammaQpSU2}. Next, we investigated the anomalous dimension of flavour
diagonal operators in $V-A$ and $V+A$ currents, given in eqs.~\eqn{gammaQmsing}
and~\eqn{gammaQpsing}. In the former, 8 penguin operators have to be added to
the operator basis while in the latter 7 additional operators are sufficient.

The anomalous dimensions of some of these operators are related to renormalon
singularities of the Borel transformed purely perturbative contribution of the
QCD correlators. More specifically, part of the dimension-6 four-quark operators
are related to the sub-leading IR singularity located at $u=3$, to which an
ambiguity in the Borel resummed series is associated. In order to investigate
the specific structure of this singularity, it is instructive to work with
diagonal bases of operators. In such bases, one isolates the combinations of
operators that do not mix under renormalisation. The anomalous dimensions are
simply the eigenvalues of the original anomalous dimension matrix. When a
four-quark operator entails an IR singularity in the Borel transformed
perturbative series, the respective eigenvalue is related to the strength of
the IR singularity at $u=3$ as given by eq.~\eqn{PoleStrength}.

The operators in the $V-A$ case do not have an associated ambiguity, since they
are order parameters of the SU$(N_f)_A$ symmetry breaking. The chiral structure
of the $V+A$ operators indicates that they have renormalon singularities
associated with them. Several singularities appear, one for each operator
combination. The most negative eigenvalue yields the strongest singularity.
In the $V+A$ non-diagonal case the corresponding exponent is $\kappa =3.173$.
In comparison with the large-$N_f$ limit of QCD, in which one has a simple and
a quadratic pole related to two dimension-6 operators, the singularity in full
QCD is stronger.

Finally, we should comment on the implications of our results to the Borel
models of the Adler function that have been used in the discussion of the RG
improvement of the perturbative series in hadronic $\tau$ decays
\cite{bj08,bbj13}. As discussed in~\cite{bbj13}, the impact of the sub-leading
IR singularity is limited since the Borel transform is dominated by the leading
IR singularity, associated with the gluon condensate. Nevertheless, the results
of the present work allow for a refined  modelling of the IR singularity at
$u=3$. However, the numerical value of the strength of the strongest IR
singularity associated with dimension-6 four-quark operators is rather close
to the one already employed in~\cite{bj08,bbj13}. The Borel transformed series
modelled in this way is not altered in any significant way. Therefore, even
taking into account the findings of the present study, the conclusions of
refs.~\cite{bj08,bbj13} remain valid.

\bigskip
\acknowledgments
Interesting discussions with Martin~Beneke, Antonio~Pineda, and Santi~Peris are
gratefully acknowledged. We also thank Martin~Beneke for carefully reading the
manuscript. DB thanks the hospitality of Universitat Aut\`onoma de Barcelona.
This work has been supported in part by the Spanish Consolider-Ingenio
2010 Programme CPAN (Grant number CSD2007-00042), by MINECO Grant numbers
CICYT-FEDER-FPA2011-25948 and CICYT-FEDER-FPA2014-55613-P, by the Severo Ochoa
excellence program of MINECO under Grant number SO-2012-0234, and by Secretaria
d'Universitats i Recerca del Departament d'Economia i Coneixement de la
Generalitat de Catalunya under Grant number 2014 SGR 1450. The work of DB
received support from the S\~ao Paulo Research Foundation (Fapesp) grant
14/50683-0.

\appendix
\section{Anomalous dimensions of four-quark operators}\label{appA}

In this Appendix, we present a generalisation of the results of reference
\cite{jk86} to an arbitrary number $N_c$ of colour degrees of freedom. In
\cite{jk86}, the leading order anomalous dimension matrix of a complete set
of local spin-zero four-quark operators without derivatives was calculated
in the case of three quark flavours.

The complete basis consists of 45 four-quark operators which in reference
\cite{jk86} were chosen as follows: with respect to the Dirac-structure, there
are five types of operators, namely, scalar, pseudoscalar, vector,
axial vector
and tensor. They can be expressed as
\begin{equation}
\bar u\Gamma u\bar d\Gamma d \,=\, \big( \bar uu\bar dd,\,
\bar u\gamma_5 u\bar d\gamma_5 d,\, \bar u\gamma_\mu u\bar d\gamma^\mu d,\,
\bar u\gamma_\mu\gamma_5 u\bar d\gamma^\mu\gamma_5 d,\,
\bar u\sigma_{\mu\nu} u\bar d\sigma^{\mu\nu} d \,\big)
\end{equation}
in the $\bar uu\bar dd$ flavour case. Employing this notation, the complete
basis $O$ of operators can be chosen to be:
\begin{eqnarray}
O \,&\equiv&\,
\big( \bar u\Gamma u\bar u\Gamma u,\, \bar d\Gamma d\bar d\Gamma d,\,
\bar s\Gamma s\bar s\Gamma s,\, \bar u\Gamma u\bar d\Gamma d,\,
\bar u\Gamma u\bar s\Gamma s,\, \bar d\Gamma d\bar s\Gamma s,\, \nn \\
\tvs
&& \hspace{3cm}
\bar u\Gamma t^a u\bar d\Gamma t^a d,\, \bar u\Gamma t^a u\bar s\Gamma t^a s,\,
\bar d\Gamma t^a d\bar s\Gamma t^a s \,\big) \,.
\end{eqnarray}

In this basis, the leading order anomalous dimension matrix takes the form
\begin{equation}
\label{gammaO1}
\gamma_O^{(1)} \,=\, \left(\! \begin{array}{ccccccccc}
A & 0 & 0 & 0 & 0 & 0 & B & B & 0 \\
0 & A & 0 & 0 & 0 & 0 & B & 0 & B \\
0 & 0 & A & 0 & 0 & 0 & 0 & B & B \\
0 & 0 & 0 & C & 0 & 0 & D & 0 & 0 \\
0 & 0 & 0 & 0 & C & 0 & 0 & D & 0 \\
0 & 0 & 0 & 0 & 0 & C & 0 & 0 & D \\
E & E & 0 & F & 0 & 0 & G & H & H \\
E & 0 & E & 0 & F & 0 & H & G & H \\
0 & E & E & 0 & 0 & F & H & H & G
\end{array} \right) \,.
\end{equation}
The submatrices are given by:
\begin{equation}
\label{Amat}
A \,=\, \left(\! \begin{array}{ccccc}
\sfrac{11}{12}-3C_F & \sfrac{7}{12} & -\,\sfrac{1}{12}+\sfrac{1}{6N_c} &
-\,\sfrac{1}{12} & -\,\sfrac{1}{8}+\sfrac{1}{4N_c} \\[2mm]
\sfrac{7}{12} & \sfrac{11}{12}-3C_F & \sfrac{1}{12}-\sfrac{1}{6N_c} &
\sfrac{1}{12} & -\,\sfrac{1}{8}+\sfrac{1}{4N_c} \\[2mm]
\sfrac{7}{6} & -\,\sfrac{7}{6} & \sfrac{11}{12}-\sfrac{1}{3N_c} &
\sfrac{11}{12}-\sfrac{3}{2N_c} & 0 \\[2mm]
-\,\sfrac{11}{6} & \sfrac{11}{6} & \sfrac{11}{12}-\sfrac{11}{6N_c} &
\sfrac{11}{12} & 0 \\[2mm]
3+\sfrac{6}{N_c} & 3+\sfrac{6}{N_c} & 0 & 0 & \sfrac{3}{2}+C_F
\end{array} \right) \,,
\end{equation}

\vspace{6mm}
\begin{equation}
\label{Bmat}
B \,=\, \left(\! \begin{array}{ccccc}
0 & 0 & -\,\sfrac{1}{3} & 0 & 0 \\[2mm]
0 & 0 & \sfrac{1}{3} & 0 & 0 \\[2mm]
0 & 0 & \sfrac{2}{3} & 0 & 0 \\[2mm]
0 & 0 & \sfrac{2}{3} & 0 & 0 \\[2mm]
0 & 0 & 0 & 0 & 0
\end{array} \right) \,,
\qquad
C \,=\, \left(\! \begin{array}{ccccc}
-\,3C_F & 0 & 0 & 0 & 0 \\[2mm]
0 & -\,3C_F & 0 & 0 & 0 \\[2mm]
0 & 0 & 0 & 0 & 0 \\[2mm]
0 & 0 & 0 & 0 & 0 \\[2mm]
0 & 0 & 0 & 0 & C_F
\end{array} \right) \,,
\end{equation}

\vspace{6mm}
\begin{equation}
\label{Dmat}
D \,=\, \left(\! \begin{array}{ccccc}
0 & 0 & 0 & 0 & -\,\sfrac{1}{2} \\[2mm]
0 & 0 & 0 & 0 & -\,\sfrac{1}{2} \\[2mm]
0 & 0 & 0 & 3 & 0 \\[2mm]
0 & 0 & 3 & 0 & 0 \\[2mm]
-12 & -12 & 0 & 0 & 0
\end{array} \right) \,,
\qquad
E \,=\, \left(\! \begin{array}{ccccc}
0 & 0 & 0 & 0 & 0 \\[2mm]
0 & 0 & 0 & 0 & 0 \\[2mm]
-\,\sfrac{1}{6} & \sfrac{1}{6} & \sfrac{1}{12}-\sfrac{1}{6N_c} &
\sfrac{1}{12} & 0 \\[2mm]
0 & 0 & 0 & 0 & 0 \\[2mm]
0 & 0 & 0 & 0 & 0
\end{array} \right) \,,
\end{equation}

\vspace{6mm}
\begin{equation}
\label{Gmat}
G \,=\, \left(\! \begin{array}{ccccc}
\sfrac{3}{2N_c} & 0 & 0 & 0 & -\,\sfrac{N_c}{8}+\sfrac{1}{2N_c} \\[2mm]
0 & \sfrac{3}{2N_c} & 0 & 0 & -\,\sfrac{N_c}{8}+\sfrac{1}{2N_c} \\[2mm]
0 & 0 & -\,\sfrac{3N_c}{4}+\sfrac{2}{3} & \sfrac{3N_c}{4}-\sfrac{3}{N_c} &
0 \\[2mm]
0 & 0 & \sfrac{3N_c}{4}-\sfrac{3}{N_c} & -\,\sfrac{3N_c}{4} & 0 \\[2mm]
-\,3N_c+\sfrac{12}{N_c} & -\,3N_c+\sfrac{12}{N_c} & 0 & 0 & C_F-\sfrac{3N_c}{2}
\end{array} \right) \,,
\end{equation}

\vspace{6mm}
\begin{equation}
\label{Hmat}
H \,=\, \left(\! \begin{array}{ccccc}
0 & 0 & 0 & 0 & 0 \\[2mm]
0 & 0 & 0 & 0 & 0 \\[2mm]
0 & 0 & \sfrac{1}{3} & 0 & 0 \\[2mm]
0 & 0 & 0 & 0 & 0 \\[2mm]
0 & 0 & 0 & 0 & 0
\end{array} \right) \,.
\end{equation}
In contrast to ref.~\cite{jk86}, here, the matrices $A$, $C$ and $G$ already
include the quark self-energy contributions depicted in figure 1c) of
\cite{jk86}, such that they are gauge independent. (The corresponding
matrices of \cite{jk86} were given in the Feynman gauge without self-energy
contribution.)

This basis of operators is particularly handy to derive the following relation
between the 10 operators of the redundant basis of non-diagonal currents given
in eqs.~\eqn{QVQAo} to \eqn{Qpmos}. We find
\begin{equation}
\label{OperatorRelation}
Q_5 \,=\, \frac{N_c}{N_c-1}\left( 2Q_+^o +2 Q_3 +2Q_4-Q_7-Q_8 \right) -
Q_+^s-Q_6+\frac{1}{2}\left(  Q_9 +Q_{10} \right).
\end{equation}

\section{Eigenvectors of anomalous dimension matrices}\label{appB}

Here, we present the eigenvectors of the non-diagonal anomalous dimension
matrices of eqs.~\eqn{gammaQm} and \eqn{gammaQp}. The coefficients of the
eigenvectors are displayed as the columns of the matrices $\hat V$ that
diagonalise the bases, as defined in eq.~\eqn{Diag}. The results are given
for $N_c=N_f=3$. For the $V-A$ case one finds
\begin{equation}
\label{evVmA}
\hat V_{Q_-} \,=\, \left(\! \begin{array}{rr}
\frac{4}{3} & -\frac{1}{6} \\[2mm]
      1     &        1
\end{array} \right) \,.
\end{equation}
The matrix $\hat V_{Q_+} $, for the $V+A$ case, reads
\begin{equation}
\label{evVpA}
\hat V_{Q_+} \,=\, \left(\! \begin{array}{rrrrrrrrr}
-0.059& -0.028&  0.472& -0.093&  0.045& -0.015&  0.316&  0.081& -0.020\\[2mm]
 0.145&  0.066& -0.664&  0.100&  0.082& -0.027&  0.949&  0.338& -0.117\\[2mm]
-0.519& -0.209& -0.254&  0.191& -0.138&  0.043&  0\po &  0.313& -0.208\\[2mm]
 0.654&  0.386& -0.159&  0.077&  0.292& -0.115&  0\po &  0.219& -0.105\\[2mm]
 0.527&  0.226&  0.496& -0.469& -0.942&  0.275&  0\po &  0.856& -0.447\\[2mm]
 0\po & -0.314&  0\po &  0.287&  0\po &  0.064&  0\po &  0\po & -0.312\\[2mm]
 0\po &  0.578&  0\po &  0.115&  0\po & -0.173&  0\po &  0\po & -0.157\\[2mm]
 0\po & -0.450&  0\po & -0.359&  0\po & -0.839&  0\po &  0\po & -0.399\\[2mm]
 0\po &  0.340&  0\po & -0.703&  0\po &  0.413&  0\po &  0\po & -0.671
\end{array} \right) \,.
\end{equation}

\bigskip
\providecommand{\href}[2]{#2}\begingroup\raggedright\endgroup

\end{document}